# The Agile Coach Role: Coaching for Agile Performance Impact

Viktoria Stray
University of Oslo, SINTEF
stray@ifi.uio.no

Anastasiia Tkalich
SINTEF
anastasiia.tkalich@sintef.no

Nils Brede Moe
SINTEF
nils.b.moe@sintef.no

## Abstract

*It is increasingly common to introduce agile coaches to help gain speed and advantage in agile companies. Following the success of Spotify, the role of the agile coach has branched out in terms of tasks and responsibilities, but little research has been conducted to examine how this role is practiced. This paper examines the role of the agile coach through 19 semi-structured interviews with agile coaches from ten different companies. We describe the role in terms of the tasks the coach has in agile projects, valuable traits, skills, tools, and the enablers of agile coaching. Our findings indicate that agile coaches perform at the team and organizational levels. They affect effort, strategies, knowledge, and skills of the agile teams. The most essential traits of an agile coach are being emphatic, people-oriented, able to listen, diplomatic, and persistent. We suggest empirically based advice for agile coaching, for example companies giving their agile coaches the authority to implement the required organizational changes within and outside the teams.*

## 1. Introduction

Agile coaching is increasingly attracting the interest of managers and software practitioners [1]. One reason is that agile coaching is among the success factors in adopting agile software development practices [2], [3]. Simultaneously, a lack of agile coaching appears to impair the development of self-managing teams [4], which is essential for agile methods.

An agile coach is often hired to help teams and companies adopt and take advantage of agile methods [5]–[7]. The role is also called a Scrum coach [8], kanban coach [9], lean coach [8], or devOps coach [10]. Agile coaches can be either hired consultants or a company's own employees who take up coaching roles [16].

The agile coach role has become especially popular following the recognition of its success at Spotify [12]. Currently, more than 58,000 people on LinkedIn have "agile coach" as a job title.[1] The success of the agile coach at Spotify can be explained by the Agile Coach Guild, a community for coaches to develop their practices and spread knowledge about agile transformation to the whole company [13]. The agile coach role at Spotify started to differentiate from that of a Scrum Master, becoming less associated with the Scrum framework and more associated with teamwork, performance, and leadership [12]. An agile coach works on increasing motivation and improving coordination, knowledge sharing, and problem solving [14].

The main goal of an agile coach is to increase teams' capabilities to attain higher levels of accomplishment and is similar to the goal of a regular team coach [15]. In addition, a recent literature review found that the main tasks of an agile coach are to support stakeholders and managers to understand and apply agile methods and facilitate and monitor effective implementation of agile throughout the whole organization [16]. However, many agile coaches are confused about their roles, tasks, and responsibilities [17].

While the role of an agile coach is popular among practitioners, the empirical knowledge about the role is still scarce. The current study seeks to fill this gap by investigating the role and its many facets. We investigate the role with its day-to-day tasks, skills, traits, and tools. Specifically, we designed a study that aimed to answer the following research questions.

RQ1—What is the role of an agile coach?

RQ2—What are enablers of agile coaching?

The remainder of this paper is organized as follows. Section 2 describes the background on agile coaching Section 3 details the research method. Section 4 reports on the results. Section 5 discusses the main findings and our study limitations. Section 6 concludes the paper and proposes future work.

[1] Searched on 29.09.20. The number was retrieved by searching "agile coach" in the "People" section of LinkedIn.

**Table 1. An overview of the interviewees**

| Country | Company | Size | Informant | Gender | Age-group | Background | Consultant | Management responsibilities |
|---|---|---|---|---|---|---|---|---|
| USA | A | 125 000 | 1 | M | 50–60 | Software tester | No | No |
| | | | 2 | M | 40–50 | Project manager | Yes | N/A |
| | | | 3 | F | 40–50 | Software tester | No | N/A |
| | | | 4 | M | 35–45 | Software developer | No | No |
| | | | 5 | F | 50–60 | Software developer | No | N/A |
| | B | 201 000 | 6 | M | 35–45 | Project manager | No | No |
| | | | 7 | M | 35–45 | Project manager | No | Yes |
| | C | 29 000 | 8 | F | 45–55 | Project manager | No | Yes |
| Norway | D | 400 | 9 | F | 35–45 | Project manager | No | Yes |
| | | | 10 | F | 35–45 | Software developer | No | Yes |
| | | | 11 | F | 45–55 | Project manager | No | Yes |
| | E | 130 | 12 | M | 40–50 | Software developer | Yes | N/A |
| | | | 13 | F | 30–40 | Software developer | Yes | N/A |
| | F | 250 | 14 | M | 45–55 | Software developer | Yes | No |
| | | | 15 | F | 40–50 | Software developer | No | Yes |
| | G | 900 | 16 | F | 35–45 | Project manager | Yes | No |
| | H | 11 000 | 17 | F | 40–50 | Software developer | Yes | No |
| | I | 250 | 18 | M | 35–45 | Software architect | Yes | Yes |
| | J | 120 | 19 | M | 35–45 | Project manager | No | No |

## 2. Background

Agile coaching improves decision-making, commitment, and team accountability [1]. Other documented benefits of agile coaching include better teamwork [18], increased understanding of agile practices, better product quality, and lower overall cost for product development [19]. A recent study suggests that teams supported by a coach improved the ability to track project progress and increased the clarity of team objectives than teams without such support [9]. There are also indications that the benefits of hiring an agile coach when adopting agile methods exceeds the financial cost [8].

Agile coaches are often regarded as team coaches [12] who work primarily with software development teams [20]. However, agile coaches have also been found to play an important role in an agile transformation [2], [21], [22]. An agile transformation often involves a change in organizational structure, cultural values, and team organization [23]. Further, research indicates that organizations experience benefits of having agile coaches working on more than just a team level. Qualitative interviews with 49 agile practitioners from 13 countries revealed that agile coaches guided managers, collaborated with HR departments, and motivated customers to work in agile ways [20]. In this work, we see an agile coach as a person that coaches teams and organizations.

One of the purposes of an agile coach is to help with agile transformation, which involves implementing and supporting self-organizing or self-managing teams. Hackman [24] argued that expert coaching is needed to foster and support such teams. Coaching can increase a team's performance and effectiveness by eliminating process loss and maximizing process gains by affecting the following aspects of teamwork [25].

1) *Effort.* Helping to build shared commitment to the group and its tasks and to minimize social loafing and motivation problems.
2) *Performance strategies.* Helping teams change habits and find novel and more appropriate ways to solve tasks.
3) *Knowledge and skill*. Foster knowledge sharing and helping avoid inappropriate weighting of members' contributions.

Additionally, organizational support is crucial for the effectiveness of self-managing units [24], [25]. Such support can be provided by organizational factors that affect a team´s effort, strategies, knowledge, and skills, for example a reward system that recognizes group effort [24].

## 3. Method

We report on a multiple-case holistic study [26], in which we studied one phenomenon in several companies in the IT sector. We chose companies that all reported having agile coaches. Therefore, the results from each case should not be regarded as objects for comparison; rather, they should be seen as complementary findings that work together to enrich our understanding of the role of agile coaches in software development organizations.

The data was collected through qualitative interviews with agile coaches from 10 companies in Norway and the United States; see details in Table 1. The companies ranged in approximate size between 120–11,000 employees in Norway and 29,000–201,000 in the United States. By the time of data collection, Company G had used agile methods for three years, while the rest of the organizations had been using agile methods for more than five years. We interviewed 19 practitioners (ten females and nine males) in the period between October and December 2019. Companies A–E were chosen because they took part in a research project on agile coaching. We further recruited through purposive and snowball non-probability sampling [27] by searching for participants based on their work as agile coaches.

"Agile coach" was the official title of most of the informants. The informants confessed that this title was not in high demand on the job market until a couple of years ago. In terms of professional background, all agile coaches could be roughly divided into two groups. Eleven informants in the first group previously worked in software teams as either developers or testers prior to becoming agile coaches. These experienced technologists acquired leadership and coaching skills due to their expertise. The second group contained eight people with backgrounds in project management or business and administration. They could be described as middle managers with technological understanding who had competences within the agile project management framework. Informants from both groups were certified in agile project management after completing some form of formal training (for example a three-day Scrum Master course).

The interviews lasted from 38 to 90 minutes, with an average length of 60 minutes. All interviews were audio-recorded and subsequently transcribed. All authors participated in the data collection, and in the majority of the interviews, two of the authors were present. The interviews were semi-structured, and we relied on an interview guide covering (1) background, (2) coaching role and purpose, (3) agile and self-managing teams, (4) typical coaching tasks and techniques, (5) enablers and challenges of agile coaching. Examples of questions we asked are as follows; *Can you describe your typical workday? What is the purpose of an agile coach? What do you do to foster self-organized teams?* and *What helps you succeed in coaching self-organized teams?*

The first and second authors analyzed the data through an inductive variant of thematic analysis [28] using NVivo 12. All instances of data that related to the role of agile coach were coded to capture their essence (e.g., reduce dependencies, help visualize the work, and hard to change people). The codes were then grouped into categories according to which aspect of agile coaching they described (e.g., tasks, personal traits, tools, or challenges). Some code groups were subsequently clustered in the overarching themes (e.g., team coaching and enablers). Analyzing the data was an iterative process, meaning that our inductive (data-driven) approach was sometimes combined with elements of a deductive approach (testing the emerging categories on the remaining data). Finally, coaching tasks were mapped to the factors effort, performance strategies, skills, and knowledge [24], [25].

## 4. Results

In this section, we first present the findings on the role of an agile coach in terms of the associated tasks, tools, skills, and traits. Then, we describe factors that are likely to enable agile coaching.

### 4.1. The tasks of an agile coach

Our results suggest that agile coaches interact with software development teams and the surrounding organizational context (e.g., employees in other departments and business stakeholders). One coach explained, "*We coach everyone from the teams and then, managers, product managers, product owners, executives, just all. Anybody the team touches, business partners and things like that"* (Informant 5). To better understand the task of an agile coach and their meaning for self-management in organizations, we mapped the tasks to the theoretical concepts that facilitate self-management and team effectiveness [24] (Table 2). We also grouped the findings according to which level they target (team vs. organization).

**Effort.** Many agile coaches increased team members´ internal motivation, thus influencing their efforts. One informant described that he gathered information on people´s personal goals to provide them with better motivation. He said, "*So what I try to do is to find out what is it that drives this person and what are things we need to talk about and focus on, that help this person become super motivated and feel there is something valuable to it*" (Informant 18).

Furthermore, the agile coaches worked on making team members feel it is "ok to fail." The goal was to promote psychological safety and encourage team members to experiment. One informant said, "*Teams need to be able to try and fail with absolutely no criticism and get praise for doing so*" (Informant 18).

Agile coaches also seek to improve teams' abilities to collaborate by drawing attention to dysfunctional interpersonal processes and suggesting ways to improve them. Informant 5 expressed, "*I use training sessions to help teams figure out how they can collaborate*

**Table 2. Effort, strategies, knowledge, skills [24], [25], and associated coaching tasks**

| Target area | Task |
| --- | --- |
| **Team** | |
| Effort | Boost internal motivation<br>Help team members feel it is ok to fail<br>Improve interpersonal interactions<br>Increase mutual trust |
| Strategies | Coach for a better work process<br>Help formulate team rules and goals<br>Formulate and follow-up action items<br>Reduce stress<br>Facilitate decision-making |
| Knowledge and skills | Help teams understand the product and its value for user and customer<br>Provide technical guidance<br>Create new teams |
| **Organization** | |
| Effort | Affect career development and reward systems |
| Strategies | Identify and reduce dependencies and bottlenecks |
| Knowledge and skills | Coach managers and business people to increase their understanding of agile<br>Coach product owners<br>Make organizational culture more adapted to agile<br>Help managers customize agile methods to their organizations |

*differently."* However, some agile coaches stated that coaching on interpersonal relationships in teams was less important. One informant expressed, "*We want teams that support their products. So, we were doing less discovery around the people in their relationships and more discovery around the product, which is where the focus should be*" (Informant 7).

The majority of the informants emphasized the role of mutual trust among the team members. One informant expressed, "*There are always things one can do to increase team´s trust. The trust is central*" (Informant 17).

On an organizational level, coaches helped modify internal career paths and reward systems to increase effort from the team members.

**Strategies**. Agile coaches make teams reflect on different aspects of their work strategies. They often coach teams for a "better team process," which is to increase process gains and eliminate process losses. Several coaches reported that they encourage teams to monitor their work through task boards (e.g., Trello), which helps control the workload and prioritize tasks. Furthermore, coaches encouraged team members to agree on team rules and discuss norms and what they could and could not expect from each other.

Informants often encouraged members to try new ways to collaborate and made sure that the team members agreed on concrete improvement measures. The suggested improvements were subsequently followed-up by agile coaches, thus helping members be consistent with their decisions. One coach stated, "*In retrospectives, I gear team members towards what I'm trying to get from the teams. And then I'll take that information and coach the teams with action items, and then follow up on these*" (Informant 4).

Agile coaches tried to help teams monitor their work processes and reduce stress associated with higher workloads. As informant 17 expressed, "*The problem is that teams do too much at the same time, so they become stressed and the quality goes down."* Just as several other informants, she encouraged her team members to work collectively on a smaller number of tasks, to speed up delivery on individual tasks. Some also coached team members to attend less meetings: *"I had to coach a tech lead to stop saying yes to attend such a high number of meetings. And I had to coach the project members to understand that they could not expect the tech lead to develop code and at the same time be involved in all discussions happening in meetings"* (Informant 10).

Agile teams are often expected to make technical decisions without external support. However, decision-making was often described as a challenge, "Facilitating *decision-making process is what I spend the most time on*" (Informant 3). Formulating the technical problem can be seen as a crucial component of the decision-making process. Our informants helped teams formulate concrete concerns by asking them open-ended questions, often during retrospectives. Informant 14 described helping team members with decisions for the developed product, "*I facilitate workshops if the team needs to discuss some issues with functionality.*"

When the agile coaches worked on the organizational level, they sought to identify and manage dependencies between the team and other units. One way to reduce dependencies was by inviting the key teams or stakeholders to common meetings or stand-ups of the coached team. Informant 14 said, "*I organized a series of meetings with the security specialists, which eventually helped my team to figure out how to payment flow should be.*" Networks could also be established across the software teams. If a team had particular challenges, the coach could recommend getting in contact with other teams that had similar challenges. Another coach said, "*My role is to detect bottlenecks, and often these are about people and common understandings. I deal with bottlenecks by facilitating dialogue between different teams*" (Informant 13).

**Knowledge and skills.** When coaching on a team level, much of the coaching was focused on teams' understanding of the products and helping members see how to maximize the product´s value for the clients and the end-users. One coach stated, "*I kept asking my team* *questions to help them figure out what it is that the business is really asking for, when they say they want this*" (Informant 1).

The coaches with technical experience were sometimes discussion partners on technical issues that arose in teams, and some reflected on technical mentoring not being in line with the agile coach's role. Informant 18 said, "*In practice, I shift between the role of coach and the role of mentor. A mentor is more instructive and can say "I would do this" when it has to do with, say, the architecture or cloud-based solutions.*" Informant 1, without a technical background, explained how she stayed away from technical discussions: "*I am aware of technology and how to apply it. But if I am becoming a technologist, then I am starting to do the team´s job instead of my job. I am starting to become a problem solver instead of facilitator."*

Many agile coaches were in charge of creating new teams, thus being able to influence members' collective skills indirectly through team composition. However, it was rarer to change how the teams were set up. Informant 1 stated that his job was not to remove team members: "*I actually had a manager asking me about whether we should terminate the guy. And I said ´My job is not that. My job is to help everybody on the team be better´. And he said ´For how long?´ And I said ´I don´t have a time limit, my job is to keep trying´."*

In addition to the team level, agile coaching was also practiced outside of the teams. Working with the team context was seen as critical for facilitating teams´ work. Informant 17 described, "*And there are some things that the teams can't do anything about themselves. Typically, processes on an organizational level. For example, how a task is financed, how teams are put together and quality processes. So, the idea is that I should focus on those processes, and specifically also on middle managers.*" Many agile coaches increased people's overall competence in agile methods through classes held for all employees interested in agile to foster agile transformation. An informant described the need: "*As agile coaches, we have the mandate to do education, coaching, mentorship, facilitation, and any type of other education or learning to help the organization move forward in our transformation effort*" (Informant 8). Some informants worked on increasing the business people's acceptance of the agile approach through informing them of agile principles, iterative software development, and the concept of self-organizing teams. Informant 3 explained, "*I often find myself coaching the business because there is a big gap between acceptance of agile mindset among the IT folks and that of the business*." Another participant said, "*When I am coaching the business, I am trying to help them and show them value of agile*" (Informant 4).

Many agile coaches scheduled individual sessions with product owners (POs) and project managers, improving their abilities to collect information about the products under development, help prioritize tasks, and facilitate the team's own planning through the agile rituals. Informant 19 described how he did this: "*I talk with a PO about the PO-role and how to best prioritize when you have four potential customers at the same time.*"

Furthermore, agile coaches often tried to transform organizational culture. If the organization is characterized by top-down control and sanctions, an agile coach may try to change that through the dialog with management. One interviewee described that she coached middle and top management in "empowering leadership." She said, "*We do leadership training to help transition from being a command-and-control leader to more of a leader that is empowering their team*" (Informant 8).

Coaches also helped external stakeholders and managers understand how agile methods best work in the context of their organizations. One agile coach pointed out that common activities are needed to form similar understanding of agile across the entire organization. She said, "*I ran workshops that we used to agree on what we mean when we say agile so that everyone has the same understanding of what agile is for us*" (Informant 17).

## 4.2 Tools and techniques

The agile coaches used a variety of tools and techniques in their daily work; see an overview in Table 3. The interviewees said it was important to know how to ask good questions and help people collaborate. One coach stated, "*It helps to know facilitation techniques to be able to facilitate people, both when they work together and in workshops*" (Informant 15). In addition, the coaches participated in meetings, such as daily stand-up meetings, to understand communication and coordination in the teams better, as well as improving the practice itself.

The coaches helped the team members in improving how they used workflow and communication tools. Informant 10 stated, "*I helped them set up weekly meetings and I taught them tricks with tags in Jira. And we also defined the workflow in terms of how we should work with testing.*"

The most frequently mentioned tool was Slack. It was commonly used for communication with team members and to improve communication in general. Informant 12 said that, as coaching was part of communication, the coaching he did was also done through Slack. Coaches also changed how teams coordinated and communicated with each other. For example, Informant 14 told us how he facilitated such a change in two teams: "*I had to help two separate teams create an understanding that they had to work with each other towards common goals and deadlines. I created a Slack channel so that the team members in both teams could have closer conversations. It really made the two teams work better together. They used to e-mail each other or talk one-on-one on Slack, but then they lost a lot of context that everyone in both teams needed. So, with the new channel, everyone saw the history and could help each other.*"

**Table 3. Tools and techniques**

| Coaching techniques | Examples |
|---|---|
| Facilitating and participating in agile ceremonies | Daily stand-up meetings, retrospective meetings, planning meetings |
| Team health checks | Comparative Agility, Atlassian health monitor |
| Facilitating digital communication channels | Slack (help create channels, and support and observe in channels) |
| Help visualize the work | Trello, Jira, value stream mapping |
| One-on-one conversations with individuals | Discuss with leaders, product owners, architects, team members, customers |
| Conduct and facilitate workshops | Team canvas, how to split tasks, how to prioritize, what is agile, how to be a good leader in an agile organization |

Many coaches used tools to facilitate discussions on roles and purpose. Informant 17 said, "*We have worked with 'Team canvas,' which is an exercise to work towards common goals and have an overview over people's roles in the team.*"

Furthermore, understanding the team's maturity in agile methods was an essential part of the coaching. One agile coach performed such monitoring every time he entered a new team. He stated, "*When we come in as coaches, we try to understand what level of agility they have as a team*" (Informant 6). The coaches also used health checks on regular teams that they coached to get metrics to understand better how they could help. Metrics were also used to understand whether team performance changed because of coaching. Informant 13 described coaching the teams through different levels, and when she achieved the "*high performing*" level, her job as a coach was not needed anymore.

## 4.3 Skills of an agile coach

One core activity of the coach is developing knowledge and skills in the teams and in the organization. In addition, coaches also need to develop their own skills and knowledge. Having knowledge about agile methods and technological understanding, communication skills, and being able to guide were described as central skills for an agile coach. The interviewees stated the importance of understanding the value, purpose, and techniques of the agile framework and said that certifications were not necessary but could sometimes improve this understanding. The majority of our respondents had years of experience from working in agile projects, which was seen by many as a prerequisite for being able to do a good job in agile coaching. "*I can do the job I do because I have years of experience; I have seen the consequences of decisions. And that, I feel, is something you don't learn in a course. You learn ceremonies and stuff like that. But knowing the effect of your decisions is impossible without experience*" (Informant 12).

Many of the coaches had technical backgrounds. The majority of the interviewees (those with and without such backgrounds) stated that having a technical background was an advantage, but not a "must-have." Some stated that a coach without technological understanding would not be able to improve the technical skills of the team and understand the team´s challenges. Informant 17 said, "*I have technical background, which helps me get respect from the developers. I feel that it actually helps understanding the technology to get them to listen to what I say*". Another interviewee said, "*There are several types of agile coaches. Some coaches teach coding practices, for example, if they work in a DevOps context. Some coaches guide the team and establish good practices to handle several aspects of the pipeline. That is, some are technical coaches, and some are more 'culture and process' types of coaches*" (Informant 13). Technological understanding and communication skills were enablers for interacting with agile teams. One coach said, "*Soft skills and hard skills come together and must work in a balanced way*" (Informant 19). Regardless of background, it was important to have the same understanding of terms. One coach explained how she had been in a dispute with another coach: *"She had another toolbox than me. Through that project I learned that also agile coaches may have conflicts if they do not speak the same language" (Informant 13).*

Some coaches stated that they had to be able to lead and guide teams. Others commented that an agile coach should not be a command-and-control leader but still needed to be able to suggest process changes and possible solutions to problems. One coach expressed, *"You are not supposed to lead or take charge. But you must make others understand what to do by asking good questions that trigger a series of thoughts by the team members that will make them able to make the change themselves*" (Informant 12). Other leadership skills were the ability to motivate and "empower" individuals and to provide operational, informational, and emotional support.

## 4.4 Traits

The essential traits reported by agile coaches were being emphatic, people-oriented, able to listen, diplomatic, and persistent, as illustrated in Figure 1.

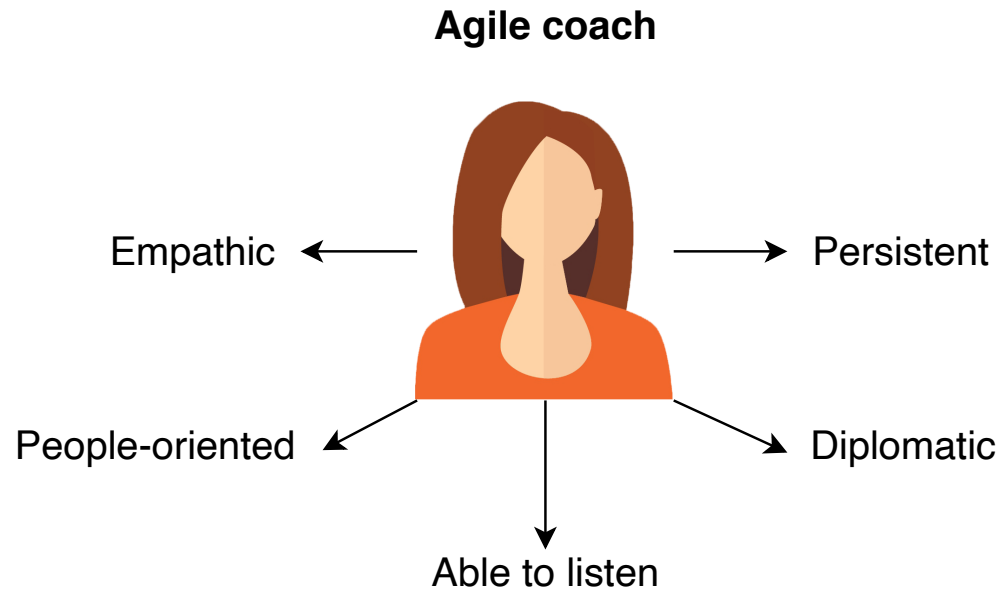

**Fig 1. Five essential traits for agile coaches**

One of the most important traits of the agile coach was **empathy**. More than half of the coaches mentioned this trait as essential for a high-performing coach because empathy helped them better understand the problems people were facing. One interviewee said, "*I use empathy a lot to understand how people are doing and to see how they react to what I say*" (Informant 17). Other coaches said they must understand different people and their perspectives and be able to do this by using their "instincts" through conversations and observations.

The majority of informants emphasized that a successful agile coach has to be **people-oriented**. Informant 15 used the phrase "*being a people-person,*" and Informant 13 said, "*It is important to wanting to get to know the people you work with.*" Informant 1 described how important it was to be present and engage with the teams. Furthermore, Informant 10 described, "*Coaching is about interactions. I like talking to people. Make them tell me how they work, and I ask questions on how they do things."* Informant 17 thought that being positive and having a sense of humor could promote better interaction at workshops: "*I use humor a lot when I conduct workshops: It is always a nice way to lower people's level of stress.*"

Next, eight informants mentioned being **able to listen**. Informant 4 expressed, "*Number one thing that I always do is I always listen. Listen to what the teams are saying and try to coach them through the bottlenecks.*" Informant 1 stated, "*I am constantly listening. I listen for the conversations that don't seem to go the right way, and I can turn and inject something.*" Informant 9 expressed, "*You must have the courage to challenge existing habits and regulations, while at the same time being humble and listen.*" Several informants indicated that bad coaches tended not to listen.

The coaches stated that being **diplomatic** was an important trait. For example, as an agile coach, one often has to give people constructive feedback, which can make them defend themselves. Informant 13 meant it is important to be diplomatic: "*You need to phrase yourself in such a way that people do not get defensive.*" In addition, it was necessary to have the ability to adapt the communication to different target groups and to explain complicated things in an accessible way.

An agile coach needs to promote change in a **persistent** manner. According to Informant 15, "*it's fun to coach, but it's also difficult, because it is not easy to change people.*" Even though people often resist changes, it is crucial for an agile coach to believe that his or her work will achieve the desired result. Informant 1 explained, "*You need to have persistence and be an optimist.*" In addition, Informant 9 highlighted, "*You have to be willing to take some battles. In a good way, of course, but still, there are some battles you have to take.*" Another interviewee said, "*You must have persistence because living in such a world is not easy, where many may not understand what you mean, and you need to explain again and again*" (Informant 15).

## 4.5 Succeeding as an agile coach

The informants described several factors helping them succeed as agile coaches. The most important was having the authority to facilitate changes. One interviewee said, "*You need to have a mandate, it doesn't help being an agile coach unless you are allowed to change things*" (Informant 12). Some coaches explained that they did not want to go into teams that were not ready to be coached: *"Some teams got the money, but they don't want to be coached yet. So, trying to coach those are always bad decisions. Over the past year, we've been able to grow evidence that allows us to not do that anymore"* (Informant 7).

The concrete tasks of agile coaches were seldom specified by organizations, which made the role flexible, but also challenging. One interviewee told us that he was confused about what the company wanted him to do, and he wished the tasks were more explicitly stated. He said, *"I found it difficult knowing what exactly to do. And, I felt that the team, because it had a lot of senior members, looked at me as I was a superfluous person."* As such, an absolute majority of our interviewees stated how important it was to have discussions with other coaches and agile practitioners. Informant 5 described that discussions with her colleague coaches provided her with professional and personal support: "*We are organized in what we call 'circles.' So, I have a circle of four other coaches and we support each other very* *closely.*" Further, some used Slack channels with coaches from different companies to discuss concrete issues and ideas on what to do as a coach.

Support from managers was also perceived as enabling, especially by agile coaches with little experience, because they had a chance to discuss with their leaders and learn from them. Besides, the top management support was described as positive for adopting agile methods in the whole organization. One coach stated, "*I think we're very lucky that agile was introduced in a top-down way*" (Informant 5). Many also stated that it was essential to identify the agile champions at the team level, which often secured agile ambassadors in teams when the coach was not present.

Finally, it was important to understand well the organization where one worked. Informant 12 said, "*You need the information from management and those outside the teams, the goal and direction, and then you can be a good discussion partner for those who work in the team*."

## 5. Discussion

We have described our findings using the concepts proposed in the theories of self-management and coaching by Hackman and Wageman [24], [25]. We will now discuss our results in relation to the research questions of the study.

### 5.1. What is the role of an agile coach?

Our results suggest that agile coaches address effort, performance strategies, knowledge, and skills. The agile coaches affected the effort of the team members by increasing their motivation to reach team goals. The coaches targeted performance strategies by coaching for better work processes. They helped team members agree on team rules and improvement measures, and they raised team members' awareness of their roles and work processes, thus helping reduce stress. Role ambiguity increases stress, and stress reduces team performance [30]. Therefore, we argue that these are important coach tasks when improving performance strategies.

The agile coaches worked on increasing knowledge and skills of the team members, by teaching new techniques and providing technical guidance. The coaches were also involved in creating new teams. As illustrated by recent multiple-case studies [23], [31], the implementation of agile methods involves a subsequent change in organizational structures, team structures, and coordination mechanisms. Our results suggest that the coaches are valuable in facilitating such changes.

The agile coach needed to have a plethora of traits. The most important were having empathy and being people-oriented. The coaches endeavored to create a safe team climate and used humor to lighten the atmosphere when coaching teams. Use of humor is in line with research on meeting behavior, which found that the use of humor and laughter triggered positive socioemotional reactions and have a long-term effect on team performance [29].

While agile coaching so far has been examined primarily with regard to software development teams [5], we found that the coaches also work with other parts of the organization. For example, they initiated discussions among managers of what agile values meant in their organizations. In this way, they made sure that organizational culture became more receptive to the agile transformation. Finally, they also coached middle and top managers in attempts to make their leadership styles more empowering and less controlling.

Furthermore, some coaches stated to increase the competence of the PO, and they facilitated the understanding of requirements between the customer and the team. We found that agile coaches play an important role in improving the quality of the developed software products and in integrating business stakeholders in the development process. An additional challenge is that business and customers may not be prepared to collaborate with agile teams [32]. Our results suggest that agile coaches address this challenge by helping the business side understand how agile software development serves their needs. A recent study found that there were discrepancies between the customers and the agile teams, and appropriate role behaviors should be agreed [32]. An agile coach can help with facilitating this agreement, as well as being proactive when tensions emerge.

### 5.2 What are enablers of agile coaching?

The personal networks of the coaches are important for their work. Our informants used their networks to receive support from other agile coaches in their own companies and externally. Discussing concrete coaching issues with colleagues at work and in agile communities was described as helpful and rewarding. Our findings confirm previous research at Spotify [13], that guilds contribute to successful agile coaching because it cultivates knowledge sharing and collective decision-making. Furthermore, our findings support previous research stating that many agile coaches are confused about their roles and responsibilities [17], thus having a network to discuss with help mitigating this challenge. In addition, a recent study found that there is occasionally tension between agile coaches and project managers because of an unclear distribution of responsibilities between these roles [33]. Therefore, enablers of agile coaching were, to more explicitly state the tasks, responsibilities, and mandates of the agile coaches in the organization.

Organizations must support the use of digital tools for team members, managers, and customers. The agile coaches frequently mentioned Slack, and they used it to communicate with team members and to improve communication in general. For example, they helped employees understand the value of openness and transparency in their communications, and they helped create channels to bridge different teams. Encouraging such communication and use of collaboration tools are in line with recent research that found Slack increased transparency in teams and facilitated network building [34].

### 5.3. Implications for practice

Our study has several implications for practice. First, it highlights the importance of the organizational context for the success of agile coaching. Without being able to affect the organizational context, agile coaches cannot assist their teams in demonstrating their full potential. Specifically, companies need to give their agile coaches the authority to implement the required changes within and outside the agile teams. For example, the coach must be able to reduce dependencies between teams, as not managing external dependencies is a barrier for self-management [35]. Organizations must also strive to make the agile coach role less ambiguous, as discussed in the previous section. Finally, educational institutions should teach agile coaching skills, such as facilitation techniques, prioritizing work, and leading teams in an empowering way.

### 5.4. Limitations

As in any empirical study, this study has some limitations. The study is qualitative, which is why the critique of qualitative methods, in general, also applies here (e.g., limited generalizability). Additionally, the sample consists of participants from two different countries, meaning that cultural factors may influence the role of agile coaches. A final limitation is that many of the informants from the United States were from the same company; however, we believe the high number of interviews and companies mitigate this limitation. We also interviewed people working in different domains, such as banking, entertainment, and transport.

## 6. Conclusions and Future Work

Today's software development is increasingly complex, and many companies adopt the agile methodology in more departments. The use of an agile coach is thus gaining popularity to help undertake an agile transformation. This study offers insight into the complex role of the agile coach and gives a detailed empirical description of how agile coaches work to improve performance in agile projects. The paper also raises awareness around the organizational impact of agile coaches, which is rarely addressed in the literature, and demonstrates how agile coaching can be used for better alignment between IT and customers. We found that the coaches work with teams, managers, and customers. They focus on improving flow and identifying and reducing dependencies. Regarding the traits favorable for successful agile coaches, our results suggest that the most important are being emphatic, people-oriented, able to listen, diplomatic, and persistent.

This paper focused on factors that enable agile coaching. Future work could investigate the inherent challenges of agile coaching, such as the issues and problems hindering effective agile coaching and the countermeasures coaches enact to overcome these challenges and make their work more effective. Furthermore, one should explore which activities, skills, and tools are less important for succeeding in coaching. For example, some informants coached teams on interpersonal relationships, whereas others meant it was less important. Future studies should also interview team members and managers to obtain an external perspective on the coaching role. Finally, it would be interesting to investigate how cultural differences and organizational size affect agile coaching and the role of the agile coach.

**Acknowledgements**

Scrum Alliance and the Research Council of Norway (grant 267704) supported this research. We are also thankful to Comparative Agility and the many participants who shared their experiences with us.

## 7. References

[1] Adkins, Coaching Agile Teams: A Companion for ScrumMasters. Addison-Wesley Professional, 2010.

[2] K. Dikert, M. Paasivaara, and C. Lassenius, "Challenges and success factors for large-scale agile transformations: A systematic literature review," Journal of Systems and Software, vol. 119, pp. 87–108, Sep. 2016, doi: 10.1016/j.jss.2016.06.013.

[3] S. C. Misra, V. Kumar, and U. Kumar, "Identifying some important success factors in adopting agile software development practices," Journal of Systems and Software, vol. 82, no. 11, pp. 1869–1890, Nov. 2009.

[4] M. Paasivaara, B. Behm, C. Lassenius, and M. Hallikainen, "Large-scale agile transformation at Ericsson: a case study," Empirical Software Engineering, vol. 23, no. 5, pp. 2550–2596, 2018, doi: 10.1007/s10664-017-9555-8.

[5] G. Bäcklander, "Doing complexity leadership theory: How agile coaches at Spotify practise enabling leadership," Creativity and Innovation Management, vol. 28, no. 1, pp. 42–60, 2019.

[6] S. Hanly, L. Waite, L. Meadows, and R. Leaton, "Agile coaching in British Telecom: Making Strawberry Jam," in

Proceedings - AGILE Conference, 2006, 2006, vol. 2006, pp. 194–202, doi: 10.1109/AGILE.2006.13.

[7] K. Silva and C. Doss, "The growth of an agile coach community at a fortune 200 company," in Proceedings - AGILE 2007, 2007, pp. 225–228, doi: 10.1109/AGILE.2007.56.

[8] R. V O'Connor and N. Duchonova, "Assessing the value of an agile coach in agile method adoption," in European Conference on Software Process Improvement, 2014, pp. 135–146.

[9] I. Shamshurin and J. S. Saltz, "Using a coach to improve team performance when the team uses a Kanban process methodology," International Journal of Information Systems and Project Management, vol. 7, no. 2, pp. 61–77, 2019.

[10] A. Wiedemann, M. Wiesche, J. B. Thatcher, and H. Gewald, "A Control-Alignment Model for Product Orientation in DevOps Teams–A Multinational Case Study," Fortieth International Conference on Information Systems (ICIS), 2019.

[11] R. Hoda, J. Noble, and M. Stuart, "Supporting self-organizing agile teams: What's senior management got to do with it.," in Agile Processes in Software Engineering and Extreme Programming, XP 2011, Madrid: Springer, 2011, pp. 73–87.

[12] G. Bäcklander, "Doing complexity leadership theory: How agile coaches at Spotify practise enabling leadership," Creativity and Innovation Management, vol. 28, no. 1, pp. 42–60, 2019, https://doi.org/10.1111/caim.12303.

[13] D. Smite, N. B. Moe, G. Levinta, and M. Floryan, "Spotify guilds: how to succeed with knowledge sharing in large-scale agile organizations," IEEE Software, vol. 36, no. 2, pp. 51–57, 2019.

[14] N. B. Moe, D. S. Cruzes, T. Dyba, and E. Engebretsen, "Coaching a Global Agile Virtual Team," Jul. 2015, pp. 33–37. IEEE 10th International Conference on Global Software Engineering, 2015.

[15] D. S. DeRue, C. M. Barnes, and F. P. Morgeson, "Understanding the motivational contingencies of team leadership," Small Group Research, vol. 41, no. 5, pp. 621–651, 2010.

[16] V. Stray, B. Memon, and L. Paruch, "A Systematic Literature Review on Agile Coaching and the Role of the Agile Coach," International Conference on Product-Focused Software Process Improvement, 2020, https://doi.org/10.1007/978-3-030-64148-1.

[17] R. Hoda, J. Noble, and S. Marshall, "Self-organizing roles on agile software development teams," IEEE Transactions on Software Engineering, vol. 39, no. 3, pp. 422–444, 2013.

[18] S. Kaltenecker and B. Myllerup, "Agile and Systemic Coaching," Scrum Alliance. https://www. scrumalliance. org/community/articles/2011/may/agile-systemic-coaching, 2011.

[19] B. Victor and N. Jacobson, "We didn't quite get it," in Agile Conference, 2009, pp. 271–274, doi: 10.1109/AGILE.2009.22.

[20] R. M. Parizi, T. J. Gandomani, and M. Z. Nafchi, "Hidden facilitators of agile transition: Agile coaches and agile champions," 2014 8th Malaysian Software Engineering Conference, MySEC 2014, pp. 246–250, 2014,

[21] M. Paasivaara, B. Behm, C. Lassenius, and M. Hallikainen, "Large-scale agile transformation at Ericsson: a case study," Empirical Software Engineering, vol. 23, no. 5, pp. 2550–2596, 2018.

[22] M. Jovanović, A. Mas, A.-L. Mesquida, and B. Lalić, "Transition of organizational roles in Agile transformation process: A grounded theory approach," Journal of Systems and Software, vol. 133, pp. 174–194, 2017,

[23] C. Fuchs, "Adapting (to) Agile Methods: Exploring the Interplay of Agile Methods and Organizational Features," presented at the 52nd Hawaii International Conference on System Sciences, Jan. 2019, doi: 10.24251/HICSS.2019.842.

[24] J. R. Hackman, "The psychology of self-management in organizations," in Psychology and work: Productivity, change, and employment, M. S. Pallack and R. O. Perloff, Eds. Washington, DC: American Psycological Association, 1986.

[25] J. R. Hackman and R. Wageman, "A theory of team coaching," Academy of Management Review, vol. 30, no. 2, pp. 269–287, 2005, doi: 10.5465/AMR.2005.16387885.

[26] R. K. Yin, Case study research and Applications: Design and Methods, 6th ed. Thousand Oaks, Calif.: SAGE publications, 2018.

[27] A. Tashakkori and C. Teddlie, Mixed methodology: Combining qualitative and quantitative approaches, vol. 46. Sage, 1998.

[28] V. Braun and V. Clarke, "Using thematic analysis in psychology," Qualitative Research in Psychology, vol. 3, no. 2, pp. 77–101, 2006.

[29] N. Lehmann-Willenbrock and J. A. Allen, "How fun are your meetings? Investigating the relationship between humor patterns in team interactions and team performance.," Journal of Applied Psychology, vol. 99, no. 6, p. 1278, 2014.

[30] J. B. Windeler, L. Maruping, and V. Venkatesh, "Technical systems development risk factors: The role of empowering leadership in lowering developers' stress," Information Systems Research, vol. 28, no. 4, pp. 775–796, 2017.

[31] V. Stray, N. B. Moe, and A. Aasheim, "Dependency Management in Large-Scale Agile: A Case Study of DevOps Teams," in Proceedings of the 52nd Hawaii International Conference on System Sciences, 2019, pp. 7007–7016, doi: 10.24251/HICSS.2019.840.

[32] L. M. Maruping and S. Matook, "The Multiplex Nature of the Customer Representative Role in Agile Information Systems Development.," MIS Quarterly, vol. 44, no. 3, 2020.

[33] G. J. Miller, "Project Management Tasks in Agile Projects: A Quantitative Study," in Proceedings of the Federated Conference on Computer Science and Information Systems, 2019, vol. 18, pp. 717–721, doi: 10.15439/2019F117.

[34] V. Stray and N. B. Moe, "Understanding coordination in global software engineering: A mixed-methods study on the use of meetings and Slack," Journal of Systems and Software, 2020, vol. 170, doi: https://doi.org/10.1016/j.jss.2020.110717.

[35] N. B. Moe, B. Dahl, V. Stray, L. S. Karlsen, and S. Schjødt-Osmo, "Team autonomy in large-scale agile," in Proceedings of the 52nd Hawaii International Conference on System Sciences, 2019, doi: 10.24251/HICSS.2019.839.